\documentclass[aps,pre,reprint,amsmath,amssymb,superscriptaddress,showpacs,floatfix]{revtex4-1}
\usepackage{graphicx}
\usepackage{dcolumn}
\usepackage{bm}
\usepackage[colorlinks, allcolors=blue]{hyperref}
\usepackage[usenames, dvipsnames]{color}

\begin{document}
\title{Anomalous transport in itinerant van der Waals ferromagnets Fe$_n$GeTe$_2$ (\emph{n}=3, 4, 5) } 

\author{Jyotirmoy Sau} 
\affiliation{Department of Condensed Matter Physics and Material Science, S. N. Bose National Centre for Basic Sciences, JD Block, Sector III, Salt Lake, Kolkata 700106, India}

\author{S. R. Hassan} 
\email{shassan@imsc.res.in}
\affiliation{The institute of mathematical sciences, CIT Campus, Tharamani, Chennai, Tamil Nadu 600113, India}
\affiliation{Homi Bhabha National Institute, Training School Complex, Anushakti Nagar, Mumbai 400094, India}

\author{Nitesh Kumar}
\email{nitesh.kumar@bose.res.in}
\affiliation{Department of Condensed Matter Physics and Material Science, S. N. Bose National Centre for Basic Sciences, JD Block, Sector III, Salt Lake, Kolkata 700106, India}

\author{Manoranjan Kumar\normalfont\textsuperscript{}}
\email{manoranjan.kumar@bose.res.in}
\thanks{The last three authors contributed equally.}
\affiliation{Department of Condensed Matter Physics and Material Science, S. N. Bose National Centre for Basic Sciences, JD Block, Sector III, Salt Lake, Kolkata 700106, India}

\begin{abstract}
 Ferromagnetic (FM) semimetals Fe$_n$GeTe$_2$(n=3, 4, 5), exhibit several symmetry-protected  band-crossing points or lines near the Fermi energy (E$_F$) and these topological properties of energy bands lead to interesting transport properties. We study these materials employing the first-principle calculations and the tight-binding Hamiltonian constructed by fitting the parameters of the first principles calculation. In the presence of spin-orbit coupling (SOC) for n=3,5 a large Berry curvature (BC) concentrated on the nodal lines is observed. The consequence of the correlation of the topological nodal line and magnetic moments on anomalous Hall conductivity (AHC) $\sigma_{xy}$ and  anomalous Nernst conductivity (ANC) $\alpha_{xy}$ have been investigated. We find $\sigma_{xy}=150$ S/cm for n=3, 295 S/cm for n=4, and 90 S/cm for n=5 at 0 K, while the
 ANC is observed as $\alpha_{xy}=0.55$ A/Km for n=3,  0.10 A/Km for n=5, and 0.80 A/Km for n=4, at the E$_F$ at room temperature. 
 Our calculated AHC values at 0 K, i.e., 150 S/cm for Fe$_3$GeTe$_2$ and 90 S/cm Fe$_5$GeTe$_2$, are consistent with the experimentally reported values. Also the experimentally reported value of ANC for Fe$_5$GeTe$_2$ is close to our calculated value at room temperature, i.e., 0.10 A/Km.
 \end{abstract}

\maketitle
\section{\label{sec1}INTRODUCTION}

Recent  experimental discovery of two and three dimensional  
 non-spatial symmetry protected topological insulating materials and their prospective applications inspired further prediction of a group of novel states protected by different symmetries, i.e., crystalline symmetry over the last few years\cite{Qi,kitaev,ryu}.  Furthermore, the concept of topological band insulator has been extended to topological semimetals (TSMs)\cite{koshi,hovrava,volovi,lau,ran,hou,kobayashi}. TSMs are described by crossing valance and conduction bands in the Brillouin zone (BZ); for example, the nodes formed by touching the conduction and valance bands at a discrete point are the Weyl/Dirac semimetals since the Dirac/Weyl equations govern their low excitation behavior\cite{hou,kobayashi,Burkov}. The recent addition is topological nodal line semimetals (TNLSMs), where bands cross each other along a line or closed-loop in BZ and, in principle, it can exist in quasi-2D\cite{lin} and 3D systems\cite{chiu}.

In TNSLMs, the degenerate crossing point of the conduction and the valence bands near E$_F$ are protected by the crystal and time-reversal symmetries (TRS). Perturbations on the Hamiltonian cannot lift the degeneracy without breaking any of its symmetries. By breaking time-reversal or spin parity symmetry, these systems may have fully gapped nodal lines or gapped into nodal points. For example, the first principles calculations demonstrate that the electronic band structure of TaAs \cite{weng,huang} in the absence of SOC exhibits two nodal lines, which are protected by mirror reflection and spin-rotation symmetries. Each nodal line gaps into three pairs of Weyl nodes in the presence of SOC. A double nodal line in SrIrO$_3$ is another example of nodal lines gapping into a pair of Dirac nodes when a mirror reflection symmetry is broken \cite{carter,fang}. These nontrivial topological energy bands lead to many exotic phenomena, such as the AHE and ANE. ANE is a thermoelectric counterpart of the AHE, both of which are associated with the BC\cite{Xiao}. Moreover, compared to AHE, which investigates the Berry curvature of the whole Fermi sea, ANE is exposed to the BC near the Fermi Energy(E$_F$). Consequently, with an enhanced Berry curvature at the nodal line near the E$_F$ , ANE  evolves as the general term of the total Nernst signal in TNLSMs.

Recently, the van der Waals (vdW) ferromagnet Fe$_3$GeTe$_2$ has been
demonstrated to be a magnetic variant of TNLSM, wherein the large BC
induced AHC has been attributed to the presence of a gapped nodal line
near E$_F$\cite{seo}.  Later related iron rich compounds in the series
Fe$_4$GeTe$_2$ and Fe$_5$GeTe$_2$ were prepared\cite{seo}.
Interestingly, both are vdW compounds and exhibit ferromagnetic
behavior. This motivated us to investigate the interplay between
magnetism and topology in the magnetic vdW materials series,
Fe$_n$GeTe$_2$ (n=3, 4, 5) in order to understand their potential
anomalous transport properties.

\begin{figure*}[t]
\centering
\includegraphics[width=1.0\linewidth]{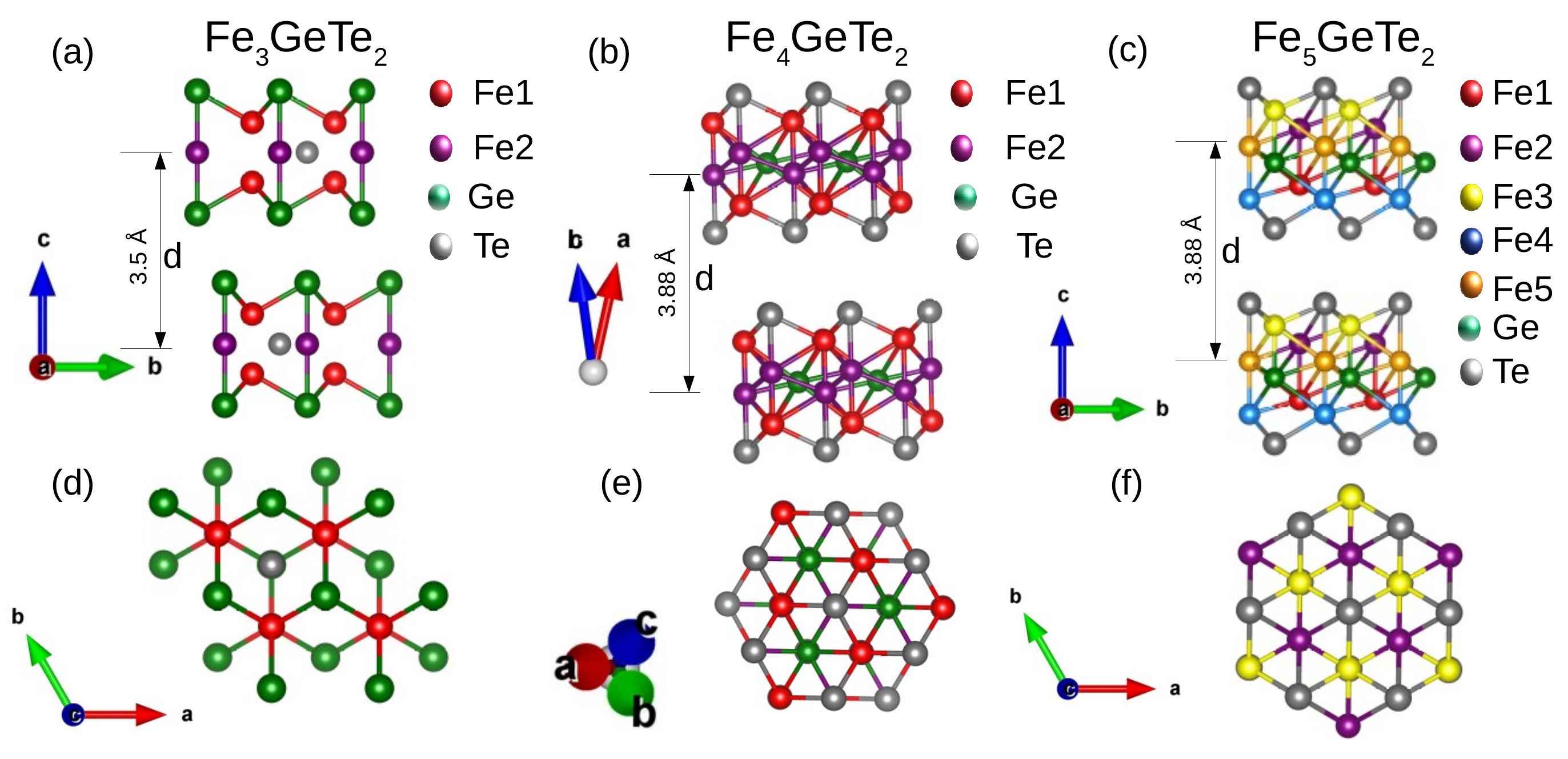}
\caption{Three stable vdW structures in the series of  Fe$_3$GeTe$_2$ for n=3, 4, 5.  The upper panels (a), (b), and (c) are side views of the structures  Fe$_3$GeTe$_2$, Fe$_4$GeTe$_2$, and Fe$_5$GeTe$_2$ respectively. While the lower panels 
 (d), (e), and (f) are corresponding top views and d represents the inter-layer distance.}
\label{fig1}
\end{figure*}

Owing to the dearth of vdW ferromagnets, Fe$_n$GeTe$_2$ have recently attracted a lot of attention. This series of compounds are also technologically important due to the large Curie temperature (T$_C$); T$_C$ = 230 K,270 K and 293 K forFe$_3$GeTe$_2$ \cite{deng,seo}, Fe$_4$GeTe$_2$\cite{suchanda} and Fe$_5$GeTe$_2$\cite{ribeiro}, respectively. Chen et al. showed that T$_C$ of Fe$_5$GeTe$_2$ could be further enhanced to 478 K upon Ni-doping\cite{chen}. The monolayer of these systems shows interesting magnetic behavior, and the thin atomic layer can be tuned using the femtosecond laser\cite{Bo}. Fe$_3$GeTe$_2$ also indicated towards high electronic correlation in terms of the emergence of Kondo behavior\cite{zhangg}. The Density functional theory study of  Fe$_3$GeTe$_2$ and  Fe$_4$GeTe$_2$ monolayers show that these systems have large magnetic anisotropy under an electric field\cite{kimm}. Fe$_5$GeTe$_2$ shows butterfly type of magneto-resistance\cite{ali}.


The structural characteristics of Fe$_n$GeTe$_2$ could be understood in terms of the triangular Fe-layers and Fe-Fe dumbbells. In Fe$_3$GeTe$_2$, Fe1 makes a triangular lattice when considered together with Ge,which can be viewed as honeycomb lattice. Fe2-Fe2 dumbbells align perpendicular to the layer, centered at each hexagon of the honeycomb, with one Fe2 above the hexagon and another Fe2 below the hexagon. Fe$_3$GeTe$_2$ belongs to space group P63/mmc (No.194). Fe$_4$GeTe$_2$ consists of two types of dumbbells each displaced with respect to the other along the direction perpendicular to the vdW layer. Fe$_4$GeTe$_2$ has a rhombohedral structure  with space group R$\bar3$m (No.166). The structure of Fe$_5$GeTe$_2$ also consists of two types of dumbbells as in the case of Fe$_4$GeTe$_2$, in addition to a triangular layer of Fe. The Space group of Fe$_5$GeTe$_2$ is P3m1 (No.156)\cite{seo}. These materials also have C$_{3y}$ symmetry in common in addition to a mirror or C$_2$ symmetry. The top and side views of Fe$_n$GeTe$_2$ structures have been depicted in Fig.\ref{fig1}.  where different Fe atoms are distinguished with separate colors.

In this manuscript, we perform a comparative study of the electronic properties of Fe$_n$GeTe$_2$ (n=3, 4, 5) to explore the origin of the different topologies in Fe$_n$GeTe$_2$. We also study AHE and ANE using a tight-binding Hamiltonian constructed by fitting parameters  with density functional theory (DFT). The magnetic moments of all these materials are close to each other; therefore, our main focus is to highlight the correlation between topology and magnetism.

The remainder of this paper is organized as follows. In  Sec.~\ref{sec2} we briefly discuss the computational details, followed by a discussion on the magnetism of these three materials in Sec.~\ref{sec3} and Sec.~\ref{sec4} and Sec.~\ref{sec5} summarise our results on the nontrivial band topology, nodal lines, and corresponding Berry curvature distribution.We present our paper's primary focus, anomalous transport, in Sec.~\ref{sec6} and conclude with a summary and conclusion in Sec.~\ref{sec7}

 \begin{figure*}[t]
\centering
\includegraphics[width=0.9\linewidth]{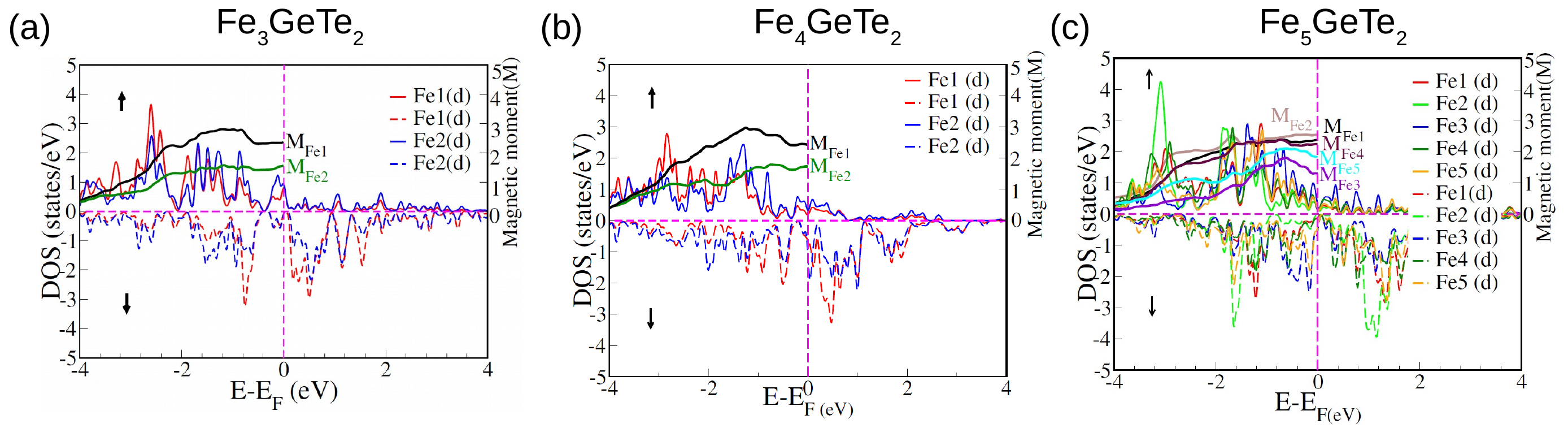}
\caption{ (a) The total DOS for Fe$_3$GeTe$_2$ as a function of (E-E$_F$) is plotted where the blue and red curves represent the total DOS for the majority spin carrier (solid) and minority spin carrier(dotted)of Fe1 and Fe2, respectively. The magnetic moments, M$_{Fe1}$ M$_{Fe2}$, are shown with the black (green) curve, the main contribution from the majority spins. (b) Similarly the total DOS for Fe$_4$GeTe$_2$ as a function of (E-E$_F$) is plotted where the blue and red curves represent the total DOS for the majority spin carrier (solid) and minority spin carrier(dotted)of Fe1 and Fe2, respectively. The magnetic moments, M$_{Fe1}$ M$_{Fe2}$, are shown with the black (green) curve. (c)the total DOS for Fe$_5$GeTe$_2$ as a function of (E-E$_F$) is plotted where the red, green, blue, deep green, orange curves represent the total DOS for the majority spin carrier (solid) and minority spin carrier (dotted) of Fe1 and Fe2, Fe3,Fe4, Fe5 respectively.}
\label{fig2}
\end{figure*}

\section{\label{sec2}Theory and Computational Details}
We examine the electronic structures and transport properties of Fe$_n$GeTe$_2$ in the scheme of DFT. First principles calculations are performed using the VASP package \cite{PE} based on the generalized gradient approximation (GGA) of the Perdew-Burke-Ernzerhof (PBE) for the exchange-correlation functional. A plane-wave basis set with a kinetic energy cutoff of 600 eV is considered while performing first-principles calculations. Furthermore, we have used the BZ sampling, a 6$\times$6$\times$1 k-point mesh forFe$_3$GeTe$_2$ and Fe$_5$GeTe$_2$, and a 6$\times$6$\times$6 k-point mesh for Fe$_4$GeTe$_2$. The Gaussian smearing method is acquired for broadening the Fermi surface with a width of 0.05 eV. Both cell parameters and internal atomic positions were fully relaxed until the forces on all atoms were smaller than 0.01 eV/A. To explore the nontrivial band topology and the intrinsic AHE, the tight-binding Hamiltonian was constructed with the maximally localized Wannier functions \cite{Pizzi_2020,marzari_97}. The intrinsic AHC is computed using the linear-response Kubo formula approach in the clean limit, and a 500$\times$500$\times$500 k-grid in the BZ was used for the integral of the AHC.

 \section{\label{sec3}Magnetic Properties}

 In this section, the magnetic properties of all three vdW compounds Fe$_3$GeTe$_2$, Fe$_4$GeTe$_2$ and Fe$_5$GeTe$_2$ are studied and we calculate spin resolved total density of states (DOS) to understand the magnetic moments. We notice that the d-orbitals of Fe atoms dictate the magnetism; therefore, we show only the DOS of the d-orbitals where  down DOS is shown as negative of its value.  We calculate the total magnetic moment of Fe atoms as a difference of total up and down spin electrons filled up to E$_F$ and density of up or down electron at any energy is proportional to the DOS of up and down spins at that energy. 
 
Fe$_3$GeTe$_2$: In this compound, there are two in-equivalent Fe atomic sites denoted as Fe1 and Fe2, as shown in Fig.\ref{fig1}(a), Fe1 lies in a-b plane and arranges on hexagonal structure, whereas  Fe2 lies perpendicular to this plane.  Energy dependence spin up (solid line) and down ( dashed line) DOS are shown in Fig.\ref{fig2}(a): red lines represent the DOS of Fe1 atoms, whereas the blue line represents DOS of Fe2 atoms. The partial DOS calculations suggest that the DOS of all d-sub-orbitals are spread near the E$_F$, therefore, electrons of all the sub-orbitals are delocalised. The DOS of Fe1 compared to Fe2 has a higher magnitude at -2.5 eV below the E$_F$, whereas the  DOS of Fe2 has a marginally larger value compared to Fe2 between the energy range 2 eV to the E$_F$.  We note that the difference between up and down DOS of Fe2 atoms have higher a spread than that of Fe1 atoms; therefore, we expect a higher magnetic moment for Fe1 than Fe2. The magnetic moments M$_{Fe1}$ and  M$_{Fe2}$ for Fe1 and Fe2 atoms are shown as solid black and green lines for Fe1 and Fe2 atoms in Fig.\ref{fig2}(a). The magnetic moment contribution of Fe1 is 
higher than that of Fe2 and these are M$_{Fe1}$=2.5 $\mu_B$/f.u. and M$_{Fe2}$=1.5 $\mu_B$/f.u. respectively at zero temperature. In Fe$_3$GeTe$_2$, the average contribution magnetic moment of Fe is M$_{Fe}$=2.12 $\mu_B$/f.u and it matches very well with experimentally reported value\cite{seo}.

\begin{figure*}[]
\centering
\includegraphics[width=1.0\linewidth]{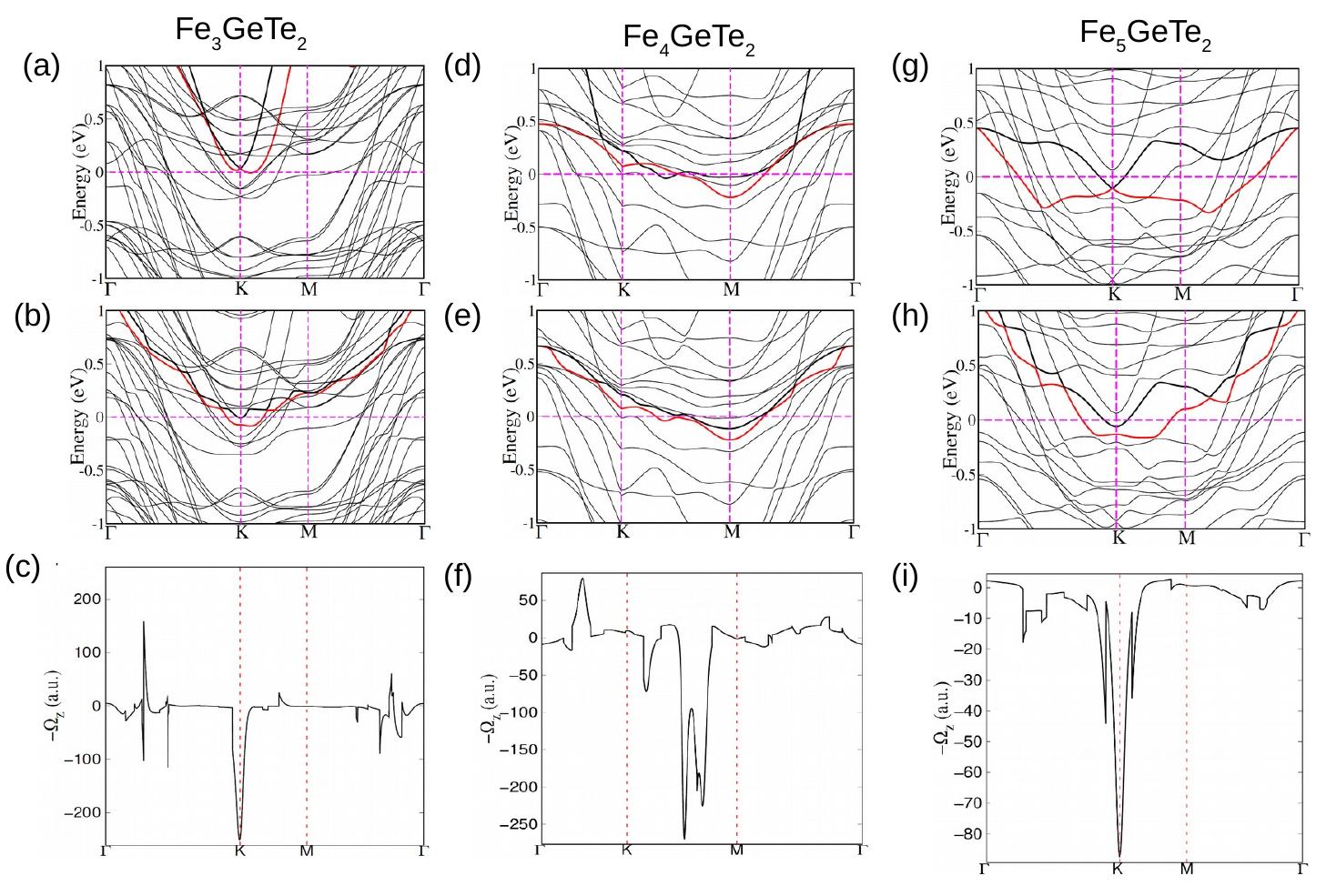}
\caption{The band structure of Fe$_n$GeTe$_2$ without and with SOC.  After opening a gap, there is a Berry curvature along symmetry lines. (a) For Fe$_3$GeTe$_2$, the crossing point at K, 0.03 eV above the Fermi energy.  The gap is opened below the Fermi energy due to SOC.  (b) The crossing point is between K and M points at the Fermi energy for Fe$_4$GeTe$_2$. (c) The crossing point at K point below the Fermi energy in Fe$_5$GeTe$_2$. With SOC, the gap opens : (d)  at the K point below the Fermi energy for Fe$_3$GeTe$_2$,  (e) between K and M point at the Fermi energy for Fe$_4$GeTe$_2$, and (f) at the K point below the Fermi energy.  A large negative Berry curvature: (g) at the K point for Fe$_3$GeTe$_2$, (h) between K and M point for Fe$_4$GeTe$_2$, and (i) at the K point for Fe$_5$GeTe$_2$.}
\label{fig3}
\end{figure*}

Fe$_4$GeTe$_2$: This system also has two in-equivalent Fe sites, labeled as Fe1 and Fe2 in Fig.\ref{fig1}(b). Fe2 atoms lying below the Te atom of hexagonal plaquettes and Fe1 atom is present in  hexagonal plaquettes. The up and down spin DOS are plotted as a function of energy with solid and dotted line. Red and blue colour represent DOS of  Fe1 and  Fe2 d-orbitals. The Fe2 $d$-bands are more dispersed compared to Fe1. The DOS of Fe1 has a larger amplitude at -2.8 eV below E$_F$ for spin up electron than Fe2 DOS, as shown in Fig.\ref{fig2}(b).  The magnetic moments M$_{Fe1}$ and  M$_{Fe2}$  for atoms Fe1 and atoms Fe2 are shown as solid black and green lines for Fe1 and Fe2 atoms in Fig.\ref{fig2}(b). Magnetic moments of  Fe1 and Fe2 are 1.7 and  2.47$\mu_B$/f.u and the average magnetic moment of Fe is M$_{Fe}$=2.084 $\mu_B$/f.u, which agrees well with the experimental value\cite{seo} . 

\begin{figure*}[]
\centering
\includegraphics[width=2.\columnwidth]{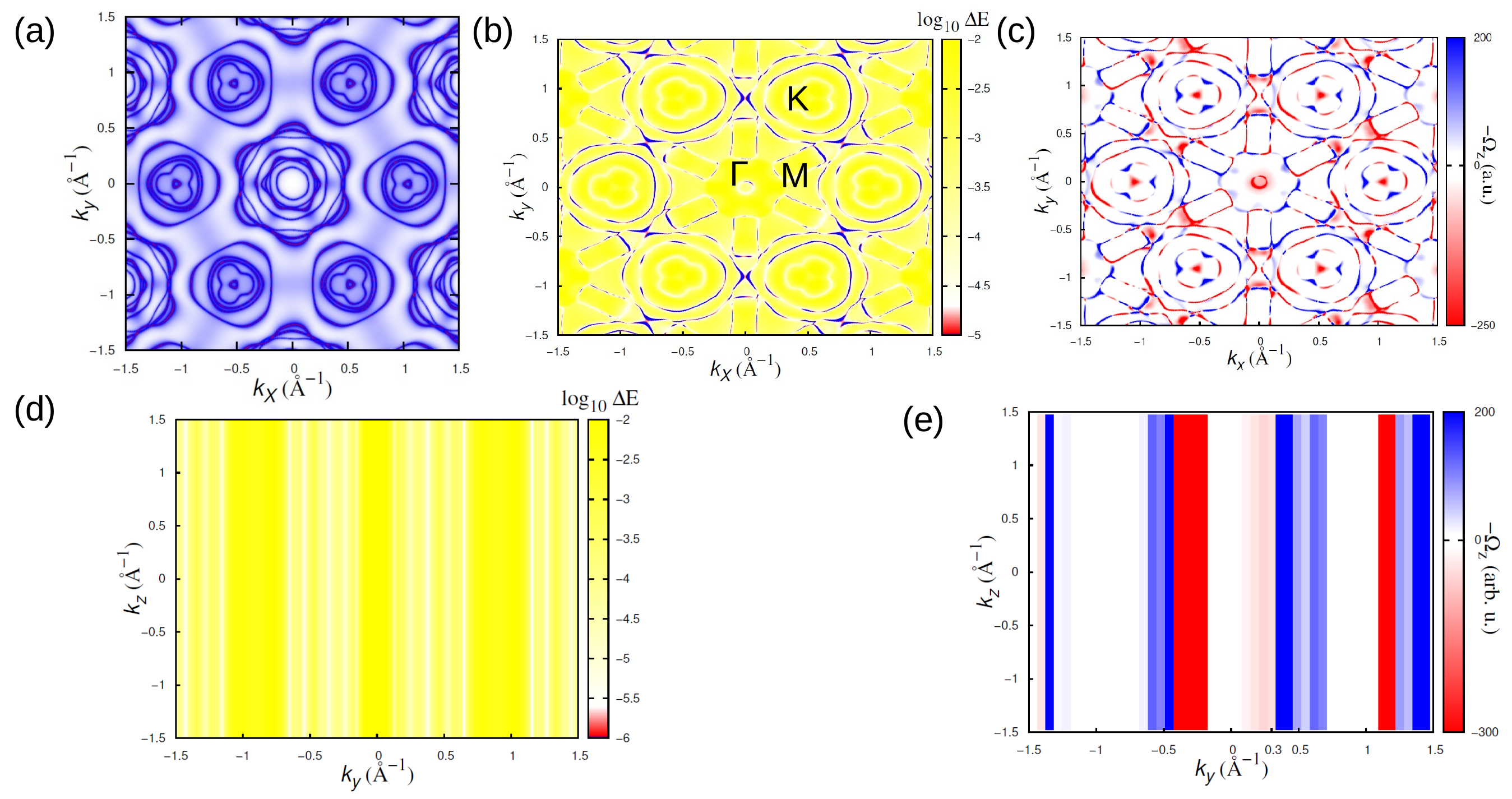}
\caption{For Fe$_3$GeTe$_2$ with SOC: (a) 2D Fermi surface using spectral function A(k$_x$,k$_y$, k$_z$=0, E$_f$). (b) Energy gap, $\Delta E(k_x,k_y,k_{z}=0) = (E_{a}(k_x,k_y,k_{z}=0)-E_{b}(k_x,k_y,k_{z}=0))$, between two crossing bands (a(red) and b(blue)) and the high symmetry points are also shown. (c) The difference in Berry curvature distribution, $\Delta \Omega_z(k_x,k_y,k_z=0)=(\Omega_{z}^{a}(k_x,k_y,k_z=0)-\Omega_{z}^{b}(k_x,k_y,k_z=0))$  of two crossing bands, has a large negative value at the K point (red spot). (d) Energy gap, $\Delta E(k_y,k_z,k_{x}=0)$, and gapped nodal  at ${k}_y$=1.2. (e) the difference of the Berry curvature distribution between two bands, $\Delta \Omega_z(k_y,k_z, k_{x}=0)$.}

\label{fig4}
\end{figure*}


Fe$_5$GeTe$_2$: It contains five in-equivalent Fe sites which are shown in Fig.\ref{fig1}(c)  and   two atoms Fe1 and Fe4 are in the same a-b plane. Fe2 and Fe3 are also in same plane and have a similar chemical environments, but have different bonding along the c-axis. d-orbitals of these atoms are also highly hybridized. The DOS of all five Fe d-orbitals are shown in Fig.\ref{fig2}(c) and DOS of Fe1 and Fe4  looks similar, whereas it is distinct for other three Fe atoms. The Green curve represents the large peak in up DOS of Fe2 at -3.0 eV below $E_F$ shows localization magnetic moment and it is contributed from d$_{z^2}$ sub-orbitals. Fe2 atoms have the highest magnetic moments and Fe5 and Fe3 have the lowest contribution.  Fe5 and Fe3  d-orbitals have higher DOS for both  up and down spins near the $E_F$.  The colored thick lines represent the magnetic moments for all five Fe atoms and the largest magnetic moment is M$_{Fe2}$=2.55$\mu_B$/f.u. for Fe2.  Magnetic moment  of Fe1 and Fe4 are  nearly the same M$_{Fe1}$=2.4$\mu_B$/f.u. and M$_{Fe4}$=2.3$\mu_B$/f.u. respectively, whereas the magnetic moment of Fe$_3$ and Fe5 are M$_{Fe3}$=1.29$\mu_B$/f.u. and M$_{Fe5}$=1.84$\mu_B$/f.u. The average magnetic moment per Fe atom is M$_{Fe}$=2.064$\mu_B$/f.u., which agrees with the experimental reported value.\cite{Li}.  
\vspace*{-4mm}
\section{\label{sec4}Nontrivial band topology}
We study the topology of energy bands of these three materials and analyze the energy band crossovers. The contributions of various sub-orbitals of Fe d-orbitals to bands near E$_F$ are also investigated.

Fe$_3$GeTe$_2$:  The energy dispersion curve near E$_F$ is shown in Fig.\ref{fig3}(a)  without SOC and in Fig.\ref{fig3}(b) with SOC for this material. As shown in Fig.3(a), in the absence of
SOC, two bands, represented by 'a' (thick red line)  and 'b' ( thick black line), cross at point K, 0.03 eV above E$_F$. These two bands are solely contributed from d orbitals of Fe1 and Fe2 atoms. These two bands are solely contributed from d orbitals  of Fe1 and  Fe2 atoms. We observe that the bands crossing resemble Mexican-hat shapes induced by the Rashba effect with quenched spins, as reported in the literature\cite{seo}. The degeneracy at the K point arises due to C$_{2y}$ symmetry in the crystal and on the application of SOC along 001 direction, lifts this degeneracy  due to breaking of the time-reversal symmetry.  We also find some other band crossings  which contribute large BC contribution which lie along the  high symmetry direction, M-$\Gamma$ and $K$-$\Gamma$, as shown in Fig.\ref{fig3}(a) . Non-trivial crossover points open a gap in the presence of the SOC as shown in Fig.\ref{fig3}(b). Therefore, we investigate the BC  (It is defined in Eqn (1) in the Sec.~\ref{sec6}.) of these bands in the above mentioned high symmetry directions. The total BC has a large magnitude corresponding to these at nontrivial band crossing  close to the E$_F$, as shown in Fig.\ref{fig3}(c). It shows the largest negative BC at the K point and relatively smaller values along the M-$\Gamma$, and $K$-$\Gamma$.



Fe$_4$GeTe$_2$:It has a rhombohedral structure, and the space group is R$\bar3$m. This material also has three-fold rotational symmetry about the z-axis (c-axis) and two-fold rotation about the y-axis. All the energy bands close to the E$_F$ are contributed by Fe atoms, Many energy bands cross close to E$_F$, as shown in Fig.\ref{fig3}(d), and these bands are formed from the d$_{xz}$ ,d$_{yz}$, and d$_{x^2-y^2}$ sub-orbitals of Fe atoms. The nontrivial degeneracies near the E$_F$ are  along the K-M and  K-$\Gamma$ symmetry points in the absence of the SOC. However, the presence of the SOC along 001 breaks the TR symmetry and lifts the degeneracy, as shown in Fig.\ref{fig3}(e). The BC in this system shows a large negative value along the K-$\Gamma$ symmetry point, as shown in Fig.\ref{fig3}(f). 

Fe$_5$GeTe$_2$: This material belongs to P3m1 symmetry group and has three-fold rotational symmetry along the z-axis (c-axis) and mirror symmetry along the y-axis. Most energy bands are contributed from the d$_{xz}$,d$_{yz}$ and d$_{x_{2}-y_{2}}$ sub-orbitals of Fe atoms. A prominent band crossing -0.11 eV below E$_F$ is detected at the K point, and the participating  bands are occupied by minority spins, as shown in Fig.\ref{fig3}(g). As was mentioned earlier, the crossing point is encircled with a red dotted circle in the absence of the SOC. There are two other crossings close to K points along the K-M and K-$\Gamma$, 0.1 eV below the E$_F$. But in the presence of the SOC, gap is opened at these points as shown in Fig.\ref{fig3}(h). The Berry curvature along the K-M and K-$\Gamma$ symmetry points are shown in Fig.\ref{fig3}(i). We notice a large negative berry curvature at the K point and two other crossings around, the K point as mentioned earlier. 

\begin{figure*}[]
\centering
\includegraphics[width=2.\columnwidth]{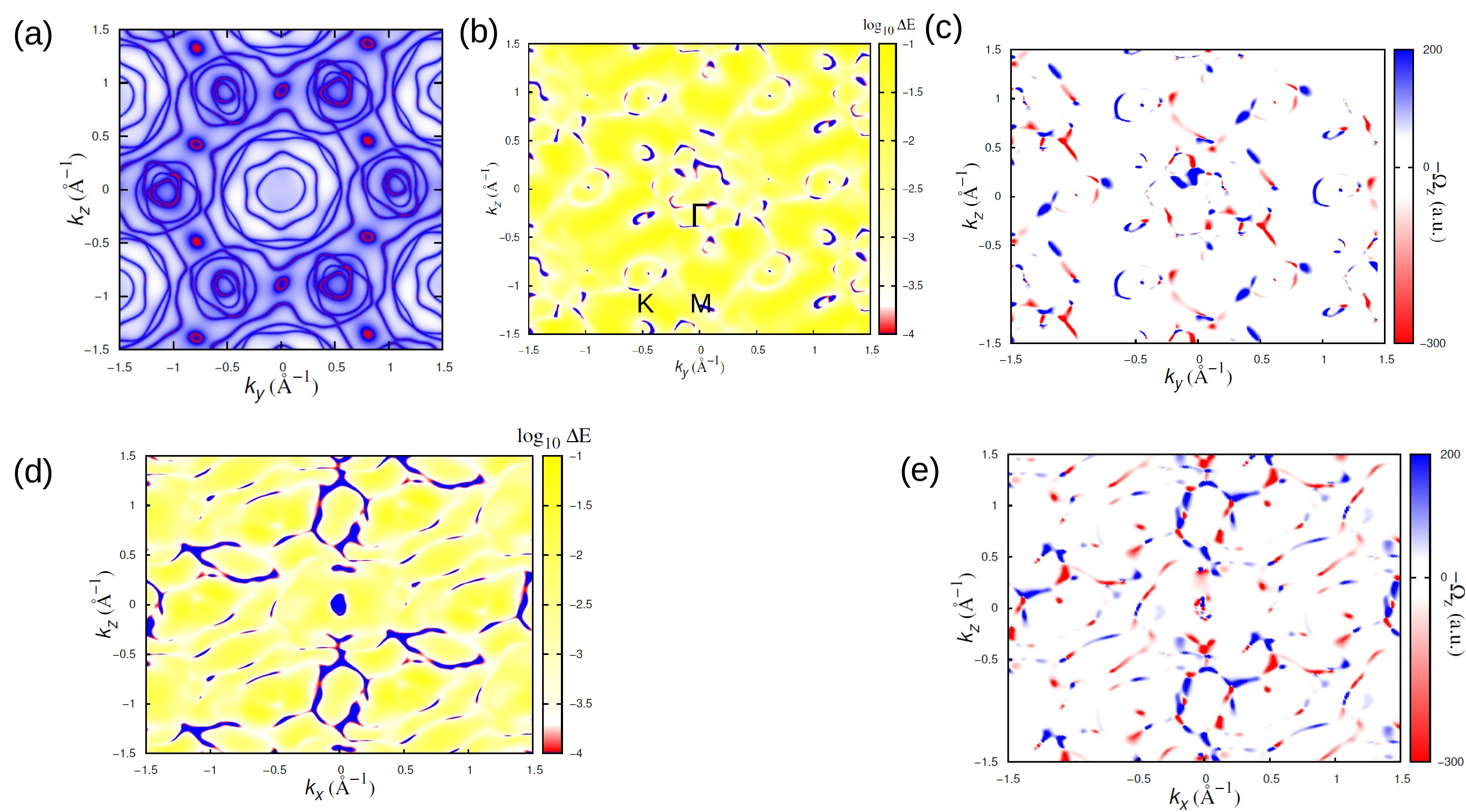}
\caption{For $Fe_5GeTe2$ with SOC: 2D Fermi surface, the energy gap, $\Delta E(k_y,k_z,k_x=0)$, and the difference of Berry curvature distribution are plotted in (a),(b), and (c), respectively.}
\label{fig5}
\end{figure*}

\section{\label{sec5}Nodal line and Berry curvature distribution}


This section highlights the band crossing feature
by the 2D Fermi surface using the spectral function A(${\bf k}$, $E_f$) computed from wanniertools\cite{WU}, the energy gap, $\Delta E$(${\bf k}$), between the  band crossing, and the difference of the z component of the Berry curvature, 
 $\Delta\Omega_z({\bf k})$,  at each point of the BZ. We usually observe large BC at the crossing points or the nodal lines. We show our results in the $k_z$=0, $k_x$=0 and $k_y$=0 planes.

Fe$_3$GeTe$_2$: Fig.\ref{fig4}(a) shows the FS plot  in $k_z=0$ plane. The calculated FS consists of a circular-shaped pocket centered at the six K-point and a hexagonal-shaped pocket centered at $\Gamma$. The red region shows a high spectral function value, and the blue has values close to zero, i.e., a small spectral function. The spectral function's intense value reveals the band's degeneracy and crossing between different bands. Most of the region is blue, i.e., low, low spectral function regime, except the red line on $\Gamma$ points and six k points exhibit high spectral function. The yellow
color represents finite gaps, whereas the white and red regions show tiny gaps opened at the high symmetry points due to the SOC. In Fig.\ref{fig4}(b), gap
distribution , $\Delta E(k_x,k_y,k_z=0)$, is shown, and a gap smaller than 10$^{-5}$ eV is considered zero. Mainly the white region is at K points, and
corresponding to this gap, there is a finite BC, as shown in
Fig.\ref{fig4}(c) with the red spot. We also studied the energy gap in
$k_x=0$ plane and observed a gapped line, as shown in Fig.\ref{fig4}(d)
by a white line (at K$_y$=1.2 \AA$^{-1}$) called a gapped nodal line.
The BC is large at this nodal line, and we show it in Fig.\ref{fig4}(e)
by a red line(at K$_y$=1.2  \AA$^{-1}$).

\begin{figure*}[t]
\centering
\includegraphics[width=2.\columnwidth]{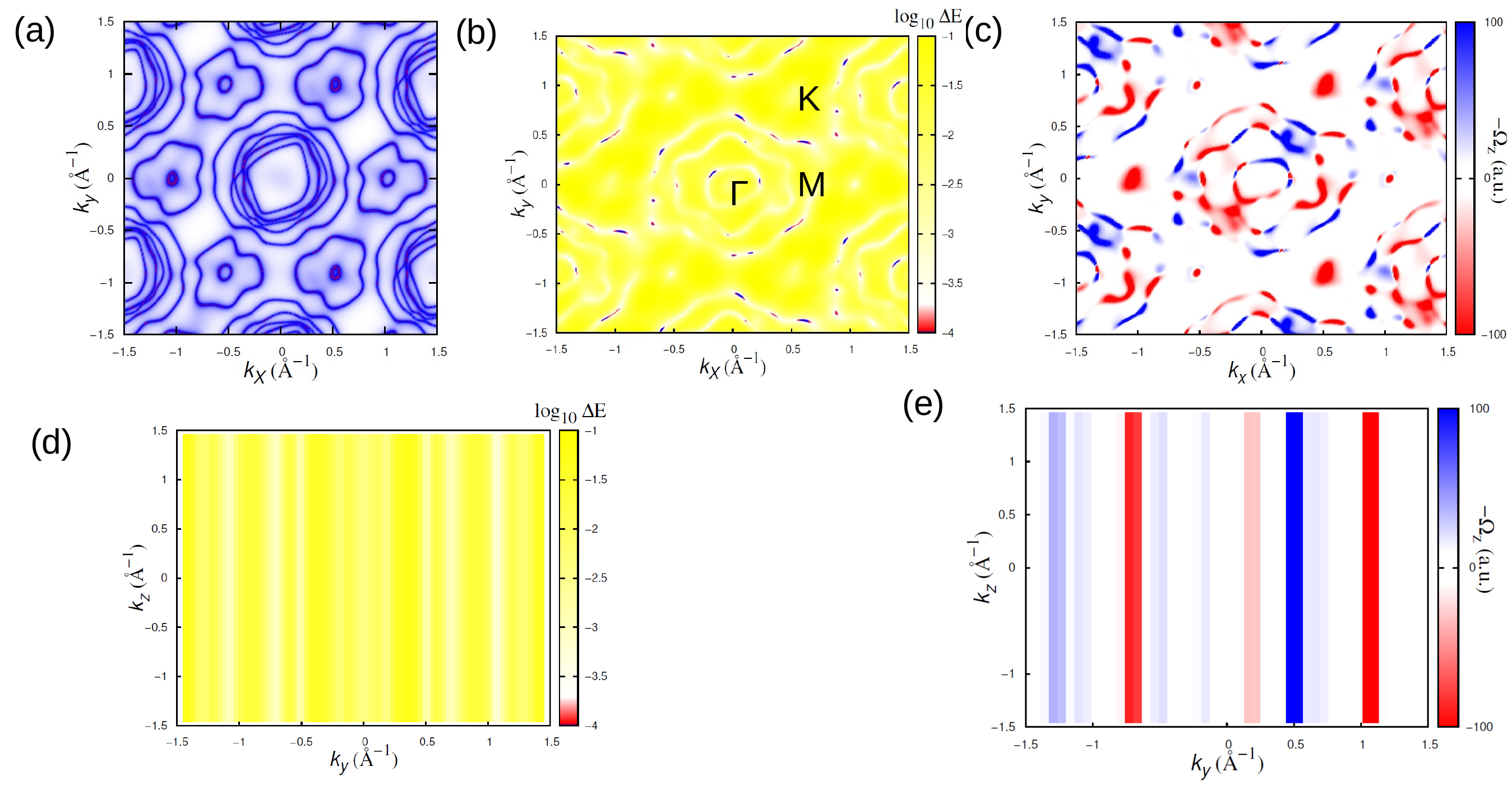}
\caption{ For Fe5 with SOC  : (a) 2D Fermi surface, (b) $\Delta E(k_x,k_y,k_z=0)$, (c)$\Delta \Omega^z(k_x,k_y,k_z=0)$, (d) $\Delta E(k_y,k_z,k_x=0)$ and (e)  $\Delta 
\Omega^z(k_y,k_z,k_x=0)$}
\label{fig6}
\end{figure*}


Fe$_4$GeTe$_2$:  In Fig.\ref{fig5}(a), 2D FS using the spectral function is plotted in the K$_x$=0 plane, and there is a $\Gamma$-centered hexagonal-shaped and six K and M point-centred circular-shaped FS. Like the previous system, the red region shows a high spectral function value indicating band degeneracy and the crossing point. The blue has values close to zero, i.e., a small spectral function. 
The intense value of the spectral function at the red dots around the $\Gamma$ points and red lines around K and M points indicate band crossing and band degeneracy. The higher symmetry around M points has consequences on the transport properties. The SOC opens tiny gaps around the degenerate points at $\Gamma$, along the K-M direction, as shown in Fig.\ref{fig5}(b). White spots between the K-M line and $\Gamma$ points correspond to gapped regions, and BC at these points are high, as shown in Fig.\ref{fig5}(c) by red spots. The gaps and BC are also calculated in the K$_y=0$ plane, but there is not much feature associated with it. Gaps and finite BC are scattered in the first BZ; therefore, we safely conclude the absence of a nodal line.

Fe$_5$GeTe$_2$: $Fe_{3}GeTe_{2}$ and $Fe_{5}GeTe_{2}$ have similar symmetries; therefore, the spectral function behavior and its symmetry are expected to be similar, as can be seen in Fig.\ref{fig6}(a) of the contour plot of FS. Most of the region is blue with a low spectral function value, except the red line with a high spectral function value in the neighborhood of $\Gamma$ points and six K points, reflecting the band degeneracy and the band crossing. In Fig.\ref{fig6}(b), gap distribution, $\Delta E(k_x,k_y)$, is shown in the first BZ, and blue
and white regions show tiny gaps opened at the high symmetry points due
to the SOC. Mainly the white region is around the K points, and 
corresponding to this gap, there is a finite BC, as shown in Fig.\ref{fig6}(c) with the red spot. In K$_x$=0 plane, white stripes represent the gaped nodal line as shown in Fig.\ref{fig6}(d), and there is a large BC associated with it as shown in Fig.\ref{fig6}(e) by a red and blue line(at K$_y$=1.2 \AA$^{-1}$).

\section{\label{sec6}Anomalous transport properties}

We also study anomalous transport properties described by a
tight-binding Hamiltonian, $H$, constructed by fitting its parameters with DFT results. As discussed in the previous section, these materials exhibit a large BC on the nodal lines in the presence of SOC. The BC (${\bf \Omega}$) acts as a magnetic monopole in the momentum space, giving rise to anomalous Hall and heat conductivities. The computation of the BC requires the energy eigenvalues $\epsilon_{n{\bf k}}$ and the energy eigenstates $\; \vert n({\bf k)} \rangle$ where ${\bf k}$ and n are the wave vector and band index respectively. ${{\bf \Omega}({\bf k})}$ is given as:

\begin{equation}
\begin{aligned}
&  {\bf\Omega^{n} ({\bf k})} =  \\ &  i\sum_{n \neq m} \frac{\langle n({\bf
k}) \vert \nabla_{\bf k}H({\bf k}) \vert m({\bf k})\rangle \times 
\langle m({\bf k}) \vert \nabla_{\bf k}H({\bf k}) \vert n({\bf k}) \rangle}{(\epsilon_{n {\bf k}}-\epsilon_{m\bf k})^2}. 
\end{aligned}
\end{equation}

The sum of the BC defines the anomalous Hall conductivity ($\sigma_{xy}$)  for all the occupied bands up to the Fermi energy E$_F$. $\sigma_{xy}$ is defined as:

\begin{equation}
\sigma_{xy} = -{\frac{e^2} {\hbar}} \sum_{n}\int\frac{d^3 k} {(2\pi)^3}
\Omega^n_z({\bf k})f_n({\bf k}) 
\end{equation}

where f$_n(\vec k)$ is the Fermi distribution function corresponding to the nth eigenstate and $\Omega_z^n({\bf k})$ is a z-component of  
$\bf {\Omega}^n ({\bf k}).$

The BC-effect also manifests in thermoelectric transport is driven by a statistical force, for example, temperature gradient. In the presence of a temperature gradient, the local current of carriers acquires an additional term from the carrier's magnetic moment in the presence of a non-uniform distribution. The extra term, an extrinsic Hall current ${\bf j}_{in}$, can be written in terms of the BC as,

\begin{eqnarray}
{\bf j}_{in}=-\frac{{\bf \nabla} T}{T} \times \frac{e}{h}\sum_{n}\int{} {d{\bf k}} \Omega^n({\bf k})[ (\epsilon_{n\bf k}-\mu)f_n({\bf k})&& \nonumber
\\+k_B T log(1+e^{\beta(\epsilon_{n\bf k}-\mu)}]
\end{eqnarray}
the Nernst conductivity $\alpha_{xy}$ can be extracted using $j_x=\alpha_{xy}(-\Delta T_y)$ and the $\alpha_{xy}$ can be written as 
 
 \begin{equation}
\alpha_{xy} = -\frac{1}{e} \int d\epsilon \frac{-df(\epsilon)}{d\mu} \sigma_{xy}(\epsilon) \frac{(\epsilon-\mu)}{T}  \end{equation}
 where intrinsic anomalous Hall conductivity $\sigma_{xy}(E_f)$ at zero temperature with Fermi-energy $E_f$. 
 
 \begin{figure}[t]
\centering
\includegraphics[width=1.0\linewidth]{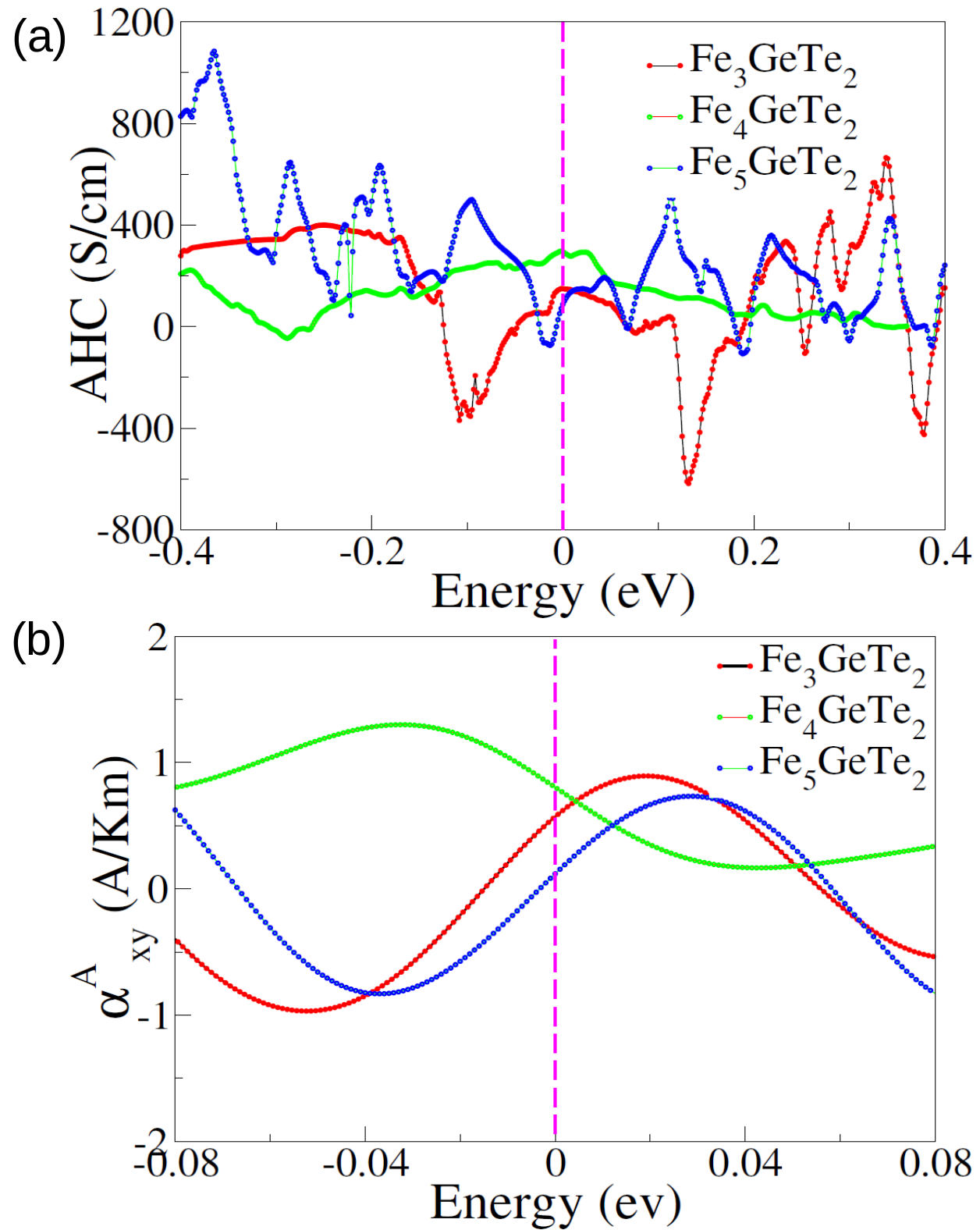}
\caption{(a) Red, green, blue curve represent  Energy ($E-E_F$) dependence of the AHC For Fe$_3$GeTe$_2$, Fe$_4$GeTe$_2$ and Fe$_5$GeTe$_2$ respectively. (b) Red, green, blue curve represent  Energy ($E-E_F$) dependence of the ANE For Fe$_3$GeTe$_2$, Fe$_4$GeTe$_2$ and Fe$_5$GeTe$_2$ respectively.} 
\label{fig7}
\end{figure}

All three materials show the variation in AHC as (E-E$_F$) varies, which is prominent in n = 3, 5 compared to n=4 and for n = 3  as the energy (E-E$_F$) varies,  AHC first decreases from 300 S/cm to -400 S/cm at (E-E$_F$) = 0.1 eV and then increases as (E-E$_F$) approaches zero and acquires a value of 150 S/cm. This material reveals the large BC at the nodal line, as seen in Fig.\ref{fig3}(c), which causes a high value of AHC. The SOC opens the gap in the energy bands below the Fermi energy, $E_F$, which contributes to a large variation in BC. Further, we can tune its values by doping, for example, up to 600 S/cm by electron doping and 450 S/cm by hole doping. Our result is consistent with the experimental value of AHC, 150 S/cm, at a low temperature. In n=4 the six-fold degeneracy around K and M points lying close to E$_F$ contributes to the large value of AHC, 295 S/cm at (E-E$_F$) = 0.4 eV decrease to -100S/cm at (E-E$_F$) = 0.3 eV and at E$_F$ its value reaches to 300 S/cm. In the case of $Fe_{5}GeTe_{2}$, the various bands contribute to BC, which is responsible for fluctuating AHC with (E-E$_F$). The calculated AHC is 90 S/cm and most of the contribution comes from the nodal line below the E$_F$.  

We now discuss ANE, which is intimately related to AHE via anomalous thermoelectric response tensor and emerges when charge carriers acquire an anomalous transverse velocity in a longitudinal temperature gradient and a finite BC. AHE probes the BC of the whole Fermi sea, while ANE is sensitive to the BC of the near Fermi energy. As a result, ANE may become the dominant term of the total Nernst signal in this series of materials with an enhanced BC near the Fermi energy. Fig.\ref{fig7}(b) shows ANE at room temperature and as a function of (E-E$_F$) for all three materials. The behavior of Fe$_3$GeTe$_2$ and Fe$_5$GeTe$_2$ are similar but opposite to Fe$_4$GeTe$_2$.  Near Fermi energy E-E$_F$ = 0, Fe$_4$GeTe$_2$ has a large value while Fe$_3$GeTe$_2$ and Fe$_5$GeTe$_2$ have the 
small value and are close to each other. This suggests that band crossing near the Fermi energy in all these materials is different, as evident in the band structure calculation discussed in the previous section. At room temperature, the anomalous Nernst conductivity is observed 0.55 A/Km for  Fe$_3$GeTe$_2$ and 0.10 A/Km for Fe$_5$GeTe$_2$. The observed Nernst conductivities at room temperature are 0.55 A/Km for Fe3GeTe2, 0.10 A/Km for Fe$_5$GeTe$_2$, and 0.8 A/Km for Fe$_5$GeTe$_2$. The experimentally observed value of ANE for Fe$_5$GeTe$_2$, 0.15 A/Km, is close to our calculated value.

\section{\label{sec7}Summary and Conclusions}

In this manuscript, we compare the electronic properties of Fe$_n$GeTe$_2$ (n=3, 4, 5) using the ab-initio theory and showed these materials are semimetal and have interesting topological electronic energy bands and finite magnetization. Fe$_3$GeTe$_2$ and Fe$_5$GeTe$_2$ have a nodal line in the electronic band, whereas the Fe$_4$GeTe$_2$ no-trivial energy band crossing in the neighborhood of the K and M high symmetry points. All these three systems have three-fold rotational symmetry about the c-axis and either two-fold mirror symmetry or rotational symmetry perpendicular to this axis and these symmetries get reflected  in spectral densities at the Fermi-surface  as shown in Fig.\ref{fig4}(a), Fig.\ref{fig5}(b) and Fig.\ref{fig6}(c). We also notice that Fe$_3$GeTe$_2$ system shows extra non-trivial degenerate points along $K-\Gamma$ and $M-\Gamma$ high symmetry line other than the bands crossing at the $K$ points found in ref. \cite{seo}.

We also show that in the presence of the SOC the time-reversal symmetry breaks and the energy degeneracies at high symmetry points get lifted. We observe the finite magnetic moments of Fe atoms which are consistent with the experimental results in the literature  \cite{seo}.  The finite magnetic moments and non-zero  BC in momentum space at these non-degenerate points lead to the anomalous Hall and Nernst conductivity. The experimentally reported AHC \cite{seo} at the low temperature is consistent with our predicted value, 150 S/cm and 90 S/cm, for Fe$_3$GeTe$_2$ and Fe$_5$GeTe$_2$, respectively. Among the three Fe$_4$GeTe$_2$ has the largest AHC, 295 S/cm,  due to the higher contribution of the BC  stemming from the near degenerate energy bands around K and M symmetry points. We obtain the ANC at the room temperature, 0.55 A/Km,  0.80 A/Km, and 0.10 A/Km, for  Fe$_3$GeTe$_2$,  Fe$_4$GeTe$_2$, and Fe$_5$GeTe$_2$, respectively. The theoretical value of ANC, 0.10 A/Km, for Fe$_5$GeTe$_2$ is close to the experimentally reported value. The large ANC in these materials offers promising applications as the new generation of thermoelectric energy conversion devices. 

In summary,  we have studied the electronic band structure of Fe$_n$GeTe$_2$ (n=3, 4, 5) materials and explored the topology in energy bands.  The magnetic moments of all three materials are close but our results showed that anomalous transport properties are not the same since the nontrivial topological points in the energy bands having large BC appearing at the different symmetry points in the BZ in each system.

\section*{Acknowledgement}

 M.K. thanks Professor Atindra Nath Pal, Professor Prabhat Mandal and Professor Thirupathaiah Setti for fruitful discussion. J.S. thanks  U.G.C  for financial support. M.K  thanks SERB for financial support through Grant Sanction No.CRG/2020/000754. N.K. thanks SERB  for financial support through Grant Sanction No.CRG/2021/002747.
 

\par

\medskip


\begin{thebibliography}{32}%
\makeatletter
\providecommand \@ifxundefined [1]{%
 \@ifx{#1\undefined}
}%
\providecommand \@ifnum [1]{%
 \ifnum #1\expandafter \@firstoftwo
 \else \expandafter \@secondoftwo
 \fi
}%
\providecommand \@ifx [1]{%
 \ifx #1\expandafter \@firstoftwo
 \else \expandafter \@secondoftwo
 \fi
}%
\providecommand \natexlab [1]{#1}%
\providecommand \enquote  [1]{``#1''}%
\providecommand \bibnamefont  [1]{#1}%
\providecommand \bibfnamefont [1]{#1}%
\providecommand \citenamefont [1]{#1}%
\providecommand \href@noop [0]{\@secondoftwo}%
\providecommand \href [0]{\begingroup \@sanitize@url \@href}%
\providecommand \@href[1]{\@@startlink{#1}\@@href}%
\providecommand \@@href[1]{\endgroup#1\@@endlink}%
\providecommand \@sanitize@url [0]{\catcode `\\12\catcode `\$12\catcode
  `\&12\catcode `\#12\catcode `\^12\catcode `\_12\catcode `\%12\relax}%
\providecommand \@@startlink[1]{}%
\providecommand \@@endlink[0]{}%
\providecommand \url  [0]{\begingroup\@sanitize@url \@url }%
\providecommand \@url [1]{\endgroup\@href {#1}{\urlprefix }}%
\providecommand \urlprefix  [0]{URL }%
\providecommand \Eprint [0]{\href }%
\providecommand \doibase [0]{http://dx.doi.org/}%
\providecommand \selectlanguage [0]{\@gobble}%
\providecommand \bibinfo  [0]{\@secondoftwo}%
\providecommand \bibfield  [0]{\@secondoftwo}%
\providecommand \translation [1]{[#1]}%
\providecommand \BibitemOpen [0]{}%
\providecommand \bibitemStop [0]{}%
\providecommand \bibitemNoStop [0]{.\EOS\space}%
\providecommand \EOS [0]{\spacefactor3000\relax}%
\providecommand \BibitemShut  [1]{\csname bibitem#1\endcsname}%
\let\auto@bib@innerbib\@empty
\bibitem [{\citenamefont {Qi}\ and\ \citenamefont {Zhang}(2011)}]{Qi}%
  \BibitemOpen
  \bibfield  {author} {\bibinfo {author} {\bibfnamefont {X.-L.}\ \bibnamefont
  {Qi}}\ and\ \bibinfo {author} {\bibfnamefont {S.-C.}\ \bibnamefont {Zhang}},\
  }\href {\doibase 10.1103/RevModPhys.83.1057} {\bibfield  {journal} {\bibinfo
  {journal} {Rev. Mod. Phys.}\ }\textbf {\bibinfo {volume} {83}},\ \bibinfo
  {pages} {1057} (\bibinfo {year} {2011})}\BibitemShut {NoStop}%
\bibitem [{\citenamefont {Kitaev}(2009)}]{kitaev}%
  \BibitemOpen
  \bibfield  {author} {\bibinfo {author} {\bibfnamefont {A.}~\bibnamefont
  {Kitaev}},\ }in\ \href@noop {} {\emph {\bibinfo {booktitle} {AIP conference
  proceedings}}},\ Vol.\ \bibinfo {volume} {1134}\ (\bibinfo {organization}
  {American Institute of Physics},\ \bibinfo {year} {2009})\ pp.\ \bibinfo
  {pages} {22--30}\BibitemShut {NoStop}%
\bibitem [{\citenamefont {Ryu}\ \emph {et~al.}(2010)\citenamefont {Ryu},
  \citenamefont {Schnyder}, \citenamefont {Furusaki},\ and\ \citenamefont
  {Ludwig}}]{ryu}%
  \BibitemOpen
  \bibfield  {author} {\bibinfo {author} {\bibfnamefont {S.}~\bibnamefont
  {Ryu}}, \bibinfo {author} {\bibfnamefont {A.~P.}\ \bibnamefont {Schnyder}},
  \bibinfo {author} {\bibfnamefont {A.}~\bibnamefont {Furusaki}}, \ and\
  \bibinfo {author} {\bibfnamefont {A.~W.}\ \bibnamefont {Ludwig}},\
  }\href@noop {} {\bibfield  {journal} {\bibinfo  {journal} {New Journal of
  Physics}\ }\textbf {\bibinfo {volume} {12}},\ \bibinfo {pages} {065010}
  (\bibinfo {year} {2010})}\BibitemShut {NoStop}%
\bibitem [{\citenamefont {Koshino}\ \emph {et~al.}(2014)\citenamefont
  {Koshino}, \citenamefont {Morimoto},\ and\ \citenamefont {Sato}}]{koshi}%
  \BibitemOpen
  \bibfield  {author} {\bibinfo {author} {\bibfnamefont {M.}~\bibnamefont
  {Koshino}}, \bibinfo {author} {\bibfnamefont {T.}~\bibnamefont {Morimoto}}, \
  and\ \bibinfo {author} {\bibfnamefont {M.}~\bibnamefont {Sato}},\ }\href@noop
  {} {\bibfield  {journal} {\bibinfo  {journal} {Physical Review B}\ }\textbf
  {\bibinfo {volume} {90}},\ \bibinfo {pages} {115207} (\bibinfo {year}
  {2014})}\BibitemShut {NoStop}%
\bibitem [{\citenamefont {Ho{\v{r}}ava}(2005)}]{hovrava}%
  \BibitemOpen
  \bibfield  {author} {\bibinfo {author} {\bibfnamefont {P.}~\bibnamefont
  {Ho{\v{r}}ava}},\ }\href@noop {} {\bibfield  {journal} {\bibinfo  {journal}
  {Physical review letters}\ }\textbf {\bibinfo {volume} {95}},\ \bibinfo
  {pages} {016405} (\bibinfo {year} {2005})}\BibitemShut {NoStop}%
\bibitem [{\citenamefont {Volovik}(2013)}]{volovi}%
  \BibitemOpen
  \bibfield  {author} {\bibinfo {author} {\bibfnamefont {G.~E.}\ \bibnamefont
  {Volovik}},\ }in\ \href@noop {} {\emph {\bibinfo {booktitle} {Analogue
  Gravity Phenomenology}}}\ (\bibinfo  {publisher} {Springer},\ \bibinfo {year}
  {2013})\ pp.\ \bibinfo {pages} {343--383}\BibitemShut {NoStop}%
\bibitem [{\citenamefont {Lau}\ and\ \citenamefont {Timm}(2013)}]{lau}%
  \BibitemOpen
  \bibfield  {author} {\bibinfo {author} {\bibfnamefont {A.}~\bibnamefont
  {Lau}}\ and\ \bibinfo {author} {\bibfnamefont {C.}~\bibnamefont {Timm}},\
  }\href@noop {} {\bibfield  {journal} {\bibinfo  {journal} {Physical Review
  B}\ }\textbf {\bibinfo {volume} {88}},\ \bibinfo {pages} {165402} (\bibinfo
  {year} {2013})}\BibitemShut {NoStop}%
\bibitem [{\citenamefont {Ran}\ \emph {et~al.}(2009)\citenamefont {Ran},
  \citenamefont {Wang}, \citenamefont {Zhai}, \citenamefont {Vishwanath},\ and\
  \citenamefont {Lee}}]{ran}%
  \BibitemOpen
  \bibfield  {author} {\bibinfo {author} {\bibfnamefont {Y.}~\bibnamefont
  {Ran}}, \bibinfo {author} {\bibfnamefont {F.}~\bibnamefont {Wang}}, \bibinfo
  {author} {\bibfnamefont {H.}~\bibnamefont {Zhai}}, \bibinfo {author}
  {\bibfnamefont {A.}~\bibnamefont {Vishwanath}}, \ and\ \bibinfo {author}
  {\bibfnamefont {D.-H.}\ \bibnamefont {Lee}},\ }\href@noop {} {\bibfield
  {journal} {\bibinfo  {journal} {Physical Review B}\ }\textbf {\bibinfo
  {volume} {79}},\ \bibinfo {pages} {014505} (\bibinfo {year}
  {2009})}\BibitemShut {NoStop}%
\bibitem [{\citenamefont {Hou}(2013)}]{hou}%
  \BibitemOpen
  \bibfield  {author} {\bibinfo {author} {\bibfnamefont {J.-M.}\ \bibnamefont
  {Hou}},\ }\href@noop {} {\bibfield  {journal} {\bibinfo  {journal} {Physical
  Review Letters}\ }\textbf {\bibinfo {volume} {111}},\ \bibinfo {pages}
  {130403} (\bibinfo {year} {2013})}\BibitemShut {NoStop}%
\bibitem [{\citenamefont {Kobayashi}\ \emph {et~al.}(2014)\citenamefont
  {Kobayashi}, \citenamefont {Shiozaki}, \citenamefont {Tanaka},\ and\
  \citenamefont {Sato}}]{kobayashi}%
  \BibitemOpen
  \bibfield  {author} {\bibinfo {author} {\bibfnamefont {S.}~\bibnamefont
  {Kobayashi}}, \bibinfo {author} {\bibfnamefont {K.}~\bibnamefont {Shiozaki}},
  \bibinfo {author} {\bibfnamefont {Y.}~\bibnamefont {Tanaka}}, \ and\ \bibinfo
  {author} {\bibfnamefont {M.}~\bibnamefont {Sato}},\ }\href@noop {} {\bibfield
   {journal} {\bibinfo  {journal} {Physical Review B}\ }\textbf {\bibinfo
  {volume} {90}},\ \bibinfo {pages} {024516} (\bibinfo {year}
  {2014})}\BibitemShut {NoStop}%
\bibitem [{\citenamefont {Burkov}\ and\ \citenamefont
  {Balents}(2011)}]{Burkov}%
  \BibitemOpen
  \bibfield  {author} {\bibinfo {author} {\bibfnamefont {A.~A.}\ \bibnamefont
  {Burkov}}\ and\ \bibinfo {author} {\bibfnamefont {L.}~\bibnamefont
  {Balents}},\ }\href {\doibase 10.1103/PhysRevLett.107.127205} {\bibfield
  {journal} {\bibinfo  {journal} {Phys. Rev. Lett.}\ }\textbf {\bibinfo
  {volume} {107}},\ \bibinfo {pages} {127205} (\bibinfo {year}
  {2011})}\BibitemShut {NoStop}%
\bibitem [{\citenamefont {Lin}\ \emph {et~al.}(2017)\citenamefont {Lin},
  \citenamefont {Hu}, \citenamefont {Chen}, \citenamefont {Lee},\ and\
  \citenamefont {Zhang}}]{lin}%
  \BibitemOpen
  \bibfield  {author} {\bibinfo {author} {\bibfnamefont {J.~Y.}\ \bibnamefont
  {Lin}}, \bibinfo {author} {\bibfnamefont {N.~C.}\ \bibnamefont {Hu}},
  \bibinfo {author} {\bibfnamefont {Y.~J.}\ \bibnamefont {Chen}}, \bibinfo
  {author} {\bibfnamefont {C.~H.}\ \bibnamefont {Lee}}, \ and\ \bibinfo
  {author} {\bibfnamefont {X.}~\bibnamefont {Zhang}},\ }\href {\doibase
  10.1103/PhysRevB.96.075438} {\bibfield  {journal} {\bibinfo  {journal} {Phys.
  Rev. B}\ }\textbf {\bibinfo {volume} {96}},\ \bibinfo {pages} {075438}
  (\bibinfo {year} {2017})}\BibitemShut {NoStop}%
\bibitem [{\citenamefont {Chiu}\ \emph {et~al.}(2013)\citenamefont {Chiu},
  \citenamefont {Yao},\ and\ \citenamefont {Ryu}}]{chiu}%
  \BibitemOpen
  \bibfield  {author} {\bibinfo {author} {\bibfnamefont {C.-K.}\ \bibnamefont
  {Chiu}}, \bibinfo {author} {\bibfnamefont {H.}~\bibnamefont {Yao}}, \ and\
  \bibinfo {author} {\bibfnamefont {S.}~\bibnamefont {Ryu}},\ }\href@noop {}
  {\bibfield  {journal} {\bibinfo  {journal} {Physical Review B}\ }\textbf
  {\bibinfo {volume} {88}},\ \bibinfo {pages} {075142} (\bibinfo {year}
  {2013})}\BibitemShut {NoStop}%
\bibitem [{\citenamefont {Weng}\ \emph {et~al.}(2015)\citenamefont {Weng},
  \citenamefont {Fang}, \citenamefont {Fang}, \citenamefont {Bernevig},\ and\
  \citenamefont {Dai}}]{weng}%
  \BibitemOpen
  \bibfield  {author} {\bibinfo {author} {\bibfnamefont {H.}~\bibnamefont
  {Weng}}, \bibinfo {author} {\bibfnamefont {C.}~\bibnamefont {Fang}}, \bibinfo
  {author} {\bibfnamefont {Z.}~\bibnamefont {Fang}}, \bibinfo {author}
  {\bibfnamefont {B.~A.}\ \bibnamefont {Bernevig}}, \ and\ \bibinfo {author}
  {\bibfnamefont {X.}~\bibnamefont {Dai}},\ }\href {\doibase
  10.1103/PhysRevX.5.011029} {\bibfield  {journal} {\bibinfo  {journal} {Phys.
  Rev. X}\ }\textbf {\bibinfo {volume} {5}},\ \bibinfo {pages} {011029}
  (\bibinfo {year} {2015})}\BibitemShut {NoStop}%
\bibitem [{\citenamefont {Huang}\ \emph {et~al.}(2015)\citenamefont {Huang},
  \citenamefont {Xu}, \citenamefont {Belopolski}, \citenamefont {Lee},
  \citenamefont {Chang}, \citenamefont {Wang}, \citenamefont {Alidoust},
  \citenamefont {Bian}, \citenamefont {Neupane}, \citenamefont {Zhang} \emph
  {et~al.}}]{huang}%
  \BibitemOpen
  \bibfield  {author} {\bibinfo {author} {\bibfnamefont {S.-M.}\ \bibnamefont
  {Huang}}, \bibinfo {author} {\bibfnamefont {S.-Y.}\ \bibnamefont {Xu}},
  \bibinfo {author} {\bibfnamefont {I.}~\bibnamefont {Belopolski}}, \bibinfo
  {author} {\bibfnamefont {C.-C.}\ \bibnamefont {Lee}}, \bibinfo {author}
  {\bibfnamefont {G.}~\bibnamefont {Chang}}, \bibinfo {author} {\bibfnamefont
  {B.}~\bibnamefont {Wang}}, \bibinfo {author} {\bibfnamefont {N.}~\bibnamefont
  {Alidoust}}, \bibinfo {author} {\bibfnamefont {G.}~\bibnamefont {Bian}},
  \bibinfo {author} {\bibfnamefont {M.}~\bibnamefont {Neupane}}, \bibinfo
  {author} {\bibfnamefont {C.}~\bibnamefont {Zhang}},  \emph {et~al.},\
  }\href@noop {} {\bibfield  {journal} {\bibinfo  {journal} {Nature
  communications}\ }\textbf {\bibinfo {volume} {6}},\ \bibinfo {pages} {1}
  (\bibinfo {year} {2015})}\BibitemShut {NoStop}%
\bibitem [{\citenamefont {Carter}\ \emph {et~al.}(2012)\citenamefont {Carter},
  \citenamefont {Shankar}, \citenamefont {Zeb},\ and\ \citenamefont
  {Kee}}]{carter}%
  \BibitemOpen
  \bibfield  {author} {\bibinfo {author} {\bibfnamefont {J.-M.}\ \bibnamefont
  {Carter}}, \bibinfo {author} {\bibfnamefont {V.~V.}\ \bibnamefont {Shankar}},
  \bibinfo {author} {\bibfnamefont {M.~A.}\ \bibnamefont {Zeb}}, \ and\
  \bibinfo {author} {\bibfnamefont {H.-Y.}\ \bibnamefont {Kee}},\ }\href
  {\doibase 10.1103/PhysRevB.85.115105} {\bibfield  {journal} {\bibinfo
  {journal} {Phys. Rev. B}\ }\textbf {\bibinfo {volume} {85}},\ \bibinfo
  {pages} {115105} (\bibinfo {year} {2012})}\BibitemShut {NoStop}%
\bibitem [{\citenamefont {Fang}\ \emph {et~al.}(2016)\citenamefont {Fang},
  \citenamefont {Lu}, \citenamefont {Liu},\ and\ \citenamefont {Fu}}]{fang}%
  \BibitemOpen
  \bibfield  {author} {\bibinfo {author} {\bibfnamefont {C.}~\bibnamefont
  {Fang}}, \bibinfo {author} {\bibfnamefont {L.}~\bibnamefont {Lu}}, \bibinfo
  {author} {\bibfnamefont {J.}~\bibnamefont {Liu}}, \ and\ \bibinfo {author}
  {\bibfnamefont {L.}~\bibnamefont {Fu}},\ }\href@noop {} {\bibfield  {journal}
  {\bibinfo  {journal} {Nature Physics}\ }\textbf {\bibinfo {volume} {12}},\
  \bibinfo {pages} {936} (\bibinfo {year} {2016})}\BibitemShut {NoStop}%
\bibitem [{\citenamefont {Xiao}\ \emph {et~al.}(2006)\citenamefont {Xiao},
  \citenamefont {Yao}, \citenamefont {Fang},\ and\ \citenamefont {Niu}}]{Xiao}%
  \BibitemOpen
  \bibfield  {author} {\bibinfo {author} {\bibfnamefont {D.}~\bibnamefont
  {Xiao}}, \bibinfo {author} {\bibfnamefont {Y.}~\bibnamefont {Yao}}, \bibinfo
  {author} {\bibfnamefont {Z.}~\bibnamefont {Fang}}, \ and\ \bibinfo {author}
  {\bibfnamefont {Q.}~\bibnamefont {Niu}},\ }\href {\doibase
  10.1103/PhysRevLett.97.026603} {\bibfield  {journal} {\bibinfo  {journal}
  {Phys. Rev. Lett.}\ }\textbf {\bibinfo {volume} {97}},\ \bibinfo {pages}
  {026603} (\bibinfo {year} {2006})}\BibitemShut {NoStop}%
\bibitem [{\citenamefont {Seo}\ \emph {et~al.}(2020)\citenamefont {Seo},
  \citenamefont {Kim}, \citenamefont {An}, \citenamefont {Kim}, \citenamefont
  {Kim}, \citenamefont {Hwang}, \citenamefont {Kim}, \citenamefont {Jang},
  \citenamefont {Kim}, \citenamefont {Eom} \emph {et~al.}}]{seo}%
  \BibitemOpen
  \bibfield  {author} {\bibinfo {author} {\bibfnamefont {J.}~\bibnamefont
  {Seo}}, \bibinfo {author} {\bibfnamefont {D.~Y.}\ \bibnamefont {Kim}},
  \bibinfo {author} {\bibfnamefont {E.~S.}\ \bibnamefont {An}}, \bibinfo
  {author} {\bibfnamefont {K.}~\bibnamefont {Kim}}, \bibinfo {author}
  {\bibfnamefont {G.-Y.}\ \bibnamefont {Kim}}, \bibinfo {author} {\bibfnamefont
  {S.-Y.}\ \bibnamefont {Hwang}}, \bibinfo {author} {\bibfnamefont {D.~W.}\
  \bibnamefont {Kim}}, \bibinfo {author} {\bibfnamefont {B.~G.}\ \bibnamefont
  {Jang}}, \bibinfo {author} {\bibfnamefont {H.}~\bibnamefont {Kim}}, \bibinfo
  {author} {\bibfnamefont {G.}~\bibnamefont {Eom}},  \emph {et~al.},\
  }\href@noop {} {\bibfield  {journal} {\bibinfo  {journal} {Science advances}\
  }\textbf {\bibinfo {volume} {6}},\ \bibinfo {pages} {eaay8912} (\bibinfo
  {year} {2020})}\BibitemShut {NoStop}%
\bibitem [{\citenamefont {Deng}\ \emph {et~al.}(2018)\citenamefont {Deng},
  \citenamefont {Yu}, \citenamefont {Song}, \citenamefont {Zhang},
  \citenamefont {Wang}, \citenamefont {Sun}, \citenamefont {Yi}, \citenamefont
  {Wu}, \citenamefont {Wu}, \citenamefont {Zhu} \emph {et~al.}}]{deng}%
  \BibitemOpen
  \bibfield  {author} {\bibinfo {author} {\bibfnamefont {Y.}~\bibnamefont
  {Deng}}, \bibinfo {author} {\bibfnamefont {Y.}~\bibnamefont {Yu}}, \bibinfo
  {author} {\bibfnamefont {Y.}~\bibnamefont {Song}}, \bibinfo {author}
  {\bibfnamefont {J.}~\bibnamefont {Zhang}}, \bibinfo {author} {\bibfnamefont
  {N.~Z.}\ \bibnamefont {Wang}}, \bibinfo {author} {\bibfnamefont
  {Z.}~\bibnamefont {Sun}}, \bibinfo {author} {\bibfnamefont {Y.}~\bibnamefont
  {Yi}}, \bibinfo {author} {\bibfnamefont {Y.~Z.}\ \bibnamefont {Wu}}, \bibinfo
  {author} {\bibfnamefont {S.}~\bibnamefont {Wu}}, \bibinfo {author}
  {\bibfnamefont {J.}~\bibnamefont {Zhu}},  \emph {et~al.},\ }\href@noop {}
  {\bibfield  {journal} {\bibinfo  {journal} {Nature}\ }\textbf {\bibinfo
  {volume} {563}},\ \bibinfo {pages} {94} (\bibinfo {year} {2018})}\BibitemShut
  {NoStop}%
\bibitem [{\citenamefont {Mondal}\ \emph {et~al.}(2021)\citenamefont {Mondal},
  \citenamefont {Khan}, \citenamefont {Mishra}, \citenamefont {Satpati},\ and\
  \citenamefont {Mandal}}]{suchanda}%
  \BibitemOpen
  \bibfield  {author} {\bibinfo {author} {\bibfnamefont {S.}~\bibnamefont
  {Mondal}}, \bibinfo {author} {\bibfnamefont {N.}~\bibnamefont {Khan}},
  \bibinfo {author} {\bibfnamefont {S.~M.}\ \bibnamefont {Mishra}}, \bibinfo
  {author} {\bibfnamefont {B.}~\bibnamefont {Satpati}}, \ and\ \bibinfo
  {author} {\bibfnamefont {P.}~\bibnamefont {Mandal}},\ }\href {\doibase
  10.1103/PhysRevB.104.094405} {\bibfield  {journal} {\bibinfo  {journal}
  {Phys. Rev. B}\ }\textbf {\bibinfo {volume} {104}},\ \bibinfo {pages}
  {094405} (\bibinfo {year} {2021})}\BibitemShut {NoStop}%
\bibitem [{\citenamefont {Ribeiro}\ \emph {et~al.}(2022)\citenamefont
  {Ribeiro}, \citenamefont {Gentile}, \citenamefont {Marty}, \citenamefont
  {Dosenovic}, \citenamefont {Okuno}, \citenamefont {Vergnaud}, \citenamefont
  {Jacquot}, \citenamefont {Jalabert}, \citenamefont {Longo}, \citenamefont
  {Ohresser} \emph {et~al.}}]{ribeiro}%
  \BibitemOpen
  \bibfield  {author} {\bibinfo {author} {\bibfnamefont {M.}~\bibnamefont
  {Ribeiro}}, \bibinfo {author} {\bibfnamefont {G.}~\bibnamefont {Gentile}},
  \bibinfo {author} {\bibfnamefont {A.}~\bibnamefont {Marty}}, \bibinfo
  {author} {\bibfnamefont {D.}~\bibnamefont {Dosenovic}}, \bibinfo {author}
  {\bibfnamefont {H.}~\bibnamefont {Okuno}}, \bibinfo {author} {\bibfnamefont
  {C.}~\bibnamefont {Vergnaud}}, \bibinfo {author} {\bibfnamefont {J.-F.}\
  \bibnamefont {Jacquot}}, \bibinfo {author} {\bibfnamefont {D.}~\bibnamefont
  {Jalabert}}, \bibinfo {author} {\bibfnamefont {D.}~\bibnamefont {Longo}},
  \bibinfo {author} {\bibfnamefont {P.}~\bibnamefont {Ohresser}},  \emph
  {et~al.},\ }\href@noop {} {\bibfield  {journal} {\bibinfo  {journal} {npj 2D
  Materials and Applications}\ }\textbf {\bibinfo {volume} {6}},\ \bibinfo
  {pages} {1} (\bibinfo {year} {2022})}\BibitemShut {NoStop}%
\bibitem [{\citenamefont {Chen}\ \emph {et~al.}(2015)\citenamefont {Chen},
  \citenamefont {Xie}, \citenamefont {Yang}, \citenamefont {Pan}, \citenamefont
  {Zhang}, \citenamefont {Cohen},\ and\ \citenamefont {Zhang}}]{chen}%
  \BibitemOpen
  \bibfield  {author} {\bibinfo {author} {\bibfnamefont {Y.}~\bibnamefont
  {Chen}}, \bibinfo {author} {\bibfnamefont {Y.}~\bibnamefont {Xie}}, \bibinfo
  {author} {\bibfnamefont {S.~A.}\ \bibnamefont {Yang}}, \bibinfo {author}
  {\bibfnamefont {H.}~\bibnamefont {Pan}}, \bibinfo {author} {\bibfnamefont
  {F.}~\bibnamefont {Zhang}}, \bibinfo {author} {\bibfnamefont {M.~L.}\
  \bibnamefont {Cohen}}, \ and\ \bibinfo {author} {\bibfnamefont
  {S.}~\bibnamefont {Zhang}},\ }\href@noop {} {\bibfield  {journal} {\bibinfo
  {journal} {Nano letters}\ }\textbf {\bibinfo {volume} {15}},\ \bibinfo
  {pages} {6974} (\bibinfo {year} {2015})}\BibitemShut {NoStop}%
\bibitem [{\citenamefont {Liu}\ \emph {et~al.}(2020)\citenamefont {Liu},
  \citenamefont {Liu}, \citenamefont {Yang}, \citenamefont {Chen},
  \citenamefont {Zhang}, \citenamefont {Li}, \citenamefont {Wu}, \citenamefont
  {Ruan}, \citenamefont {Xiu}, \citenamefont {Liu}, \citenamefont {He},
  \citenamefont {Zhang},\ and\ \citenamefont {Xu}}]{Bo}%
  \BibitemOpen
  \bibfield  {author} {\bibinfo {author} {\bibfnamefont {B.}~\bibnamefont
  {Liu}}, \bibinfo {author} {\bibfnamefont {S.}~\bibnamefont {Liu}}, \bibinfo
  {author} {\bibfnamefont {L.}~\bibnamefont {Yang}}, \bibinfo {author}
  {\bibfnamefont {Z.}~\bibnamefont {Chen}}, \bibinfo {author} {\bibfnamefont
  {E.}~\bibnamefont {Zhang}}, \bibinfo {author} {\bibfnamefont
  {Z.}~\bibnamefont {Li}}, \bibinfo {author} {\bibfnamefont {J.}~\bibnamefont
  {Wu}}, \bibinfo {author} {\bibfnamefont {X.}~\bibnamefont {Ruan}}, \bibinfo
  {author} {\bibfnamefont {F.}~\bibnamefont {Xiu}}, \bibinfo {author}
  {\bibfnamefont {W.}~\bibnamefont {Liu}}, \bibinfo {author} {\bibfnamefont
  {L.}~\bibnamefont {He}}, \bibinfo {author} {\bibfnamefont {R.}~\bibnamefont
  {Zhang}}, \ and\ \bibinfo {author} {\bibfnamefont {Y.}~\bibnamefont {Xu}},\
  }\href {\doibase 10.1103/PhysRevLett.125.267205} {\bibfield  {journal}
  {\bibinfo  {journal} {Phys. Rev. Lett.}\ }\textbf {\bibinfo {volume} {125}},\
  \bibinfo {pages} {267205} (\bibinfo {year} {2020})}\BibitemShut {NoStop}%
\bibitem [{\citenamefont {Zhang}\ \emph {et~al.}(2018)\citenamefont {Zhang},
  \citenamefont {Lu}, \citenamefont {Zhu}, \citenamefont {Tan}, \citenamefont
  {Feng}, \citenamefont {Liu}, \citenamefont {Zhang}, \citenamefont {Chen},
  \citenamefont {Liu}, \citenamefont {Luo} \emph {et~al.}}]{zhangg}%
  \BibitemOpen
  \bibfield  {author} {\bibinfo {author} {\bibfnamefont {Y.}~\bibnamefont
  {Zhang}}, \bibinfo {author} {\bibfnamefont {H.}~\bibnamefont {Lu}}, \bibinfo
  {author} {\bibfnamefont {X.}~\bibnamefont {Zhu}}, \bibinfo {author}
  {\bibfnamefont {S.}~\bibnamefont {Tan}}, \bibinfo {author} {\bibfnamefont
  {W.}~\bibnamefont {Feng}}, \bibinfo {author} {\bibfnamefont {Q.}~\bibnamefont
  {Liu}}, \bibinfo {author} {\bibfnamefont {W.}~\bibnamefont {Zhang}}, \bibinfo
  {author} {\bibfnamefont {Q.}~\bibnamefont {Chen}}, \bibinfo {author}
  {\bibfnamefont {Y.}~\bibnamefont {Liu}}, \bibinfo {author} {\bibfnamefont
  {X.}~\bibnamefont {Luo}},  \emph {et~al.},\ }\href@noop {} {\bibfield
  {journal} {\bibinfo  {journal} {Science advances}\ }\textbf {\bibinfo
  {volume} {4}},\ \bibinfo {pages} {eaao6791} (\bibinfo {year}
  {2018})}\BibitemShut {NoStop}%
\bibitem [{\citenamefont {Kim}\ \emph {et~al.}(2021)\citenamefont {Kim},
  \citenamefont {Lee}, \citenamefont {Jang}, \citenamefont {Kim},\ and\
  \citenamefont {Shim}}]{kimm}%
  \BibitemOpen
  \bibfield  {author} {\bibinfo {author} {\bibfnamefont {D.}~\bibnamefont
  {Kim}}, \bibinfo {author} {\bibfnamefont {C.}~\bibnamefont {Lee}}, \bibinfo
  {author} {\bibfnamefont {B.~G.}\ \bibnamefont {Jang}}, \bibinfo {author}
  {\bibfnamefont {K.}~\bibnamefont {Kim}}, \ and\ \bibinfo {author}
  {\bibfnamefont {J.~H.}\ \bibnamefont {Shim}},\ }\href@noop {} {\bibfield
  {journal} {\bibinfo  {journal} {Scientific Reports}\ }\textbf {\bibinfo
  {volume} {11}},\ \bibinfo {pages} {1} (\bibinfo {year} {2021})}\BibitemShut
  {NoStop}%
\bibitem [{\citenamefont {Ali}\ \emph {et~al.}(2016)\citenamefont {Ali},
  \citenamefont {Schoop}, \citenamefont {Garg}, \citenamefont {Lippmann},
  \citenamefont {Lara}, \citenamefont {Lotsch},\ and\ \citenamefont
  {Parkin}}]{ali}%
  \BibitemOpen
  \bibfield  {author} {\bibinfo {author} {\bibfnamefont {M.~N.}\ \bibnamefont
  {Ali}}, \bibinfo {author} {\bibfnamefont {L.~M.}\ \bibnamefont {Schoop}},
  \bibinfo {author} {\bibfnamefont {C.}~\bibnamefont {Garg}}, \bibinfo {author}
  {\bibfnamefont {J.~M.}\ \bibnamefont {Lippmann}}, \bibinfo {author}
  {\bibfnamefont {E.}~\bibnamefont {Lara}}, \bibinfo {author} {\bibfnamefont
  {B.}~\bibnamefont {Lotsch}}, \ and\ \bibinfo {author} {\bibfnamefont {S.~S.}\
  \bibnamefont {Parkin}},\ }\href@noop {} {\bibfield  {journal} {\bibinfo
  {journal} {Science advances}\ }\textbf {\bibinfo {volume} {2}},\ \bibinfo
  {pages} {e1601742} (\bibinfo {year} {2016})}\BibitemShut {NoStop}%
\bibitem [{\citenamefont {Bl\"ochl}(1994)}]{PE}%
  \BibitemOpen
  \bibfield  {author} {\bibinfo {author} {\bibfnamefont {P.}~\bibnamefont
  {Bl\"ochl}},\ }\href {\doibase 10.1103/PhysRevB.50.17953} {\bibfield
  {journal} {\bibinfo  {journal} {Phys. Rev. B}\ }\textbf {\bibinfo {volume}
  {50}},\ \bibinfo {pages} {17953} (\bibinfo {year} {1994})}\BibitemShut
  {NoStop}%
\bibitem [{\citenamefont {Pizzi}\ \emph {et~al.}(2020)\citenamefont {Pizzi},
  \citenamefont {Vitale}, \citenamefont {Arita}, \citenamefont {Blügel},
  \citenamefont {Freimuth}, \citenamefont {G{\'{e}}ranton}, \citenamefont
  {Gibertini}, \citenamefont {Gresch}, \citenamefont {Johnson}, \citenamefont
  {Koretsune}, \citenamefont {Iba{\~{n}}ez-Azpiroz}, \citenamefont {Lee},
  \citenamefont {Lihm}, \citenamefont {Marchand}, \citenamefont {Marrazzo},
  \citenamefont {Mokrousov}, \citenamefont {Mustafa}, \citenamefont {Nohara},
  \citenamefont {Nomura}, \citenamefont {Paulatto}, \citenamefont
  {Ponc{\'{e}}}, \citenamefont {Ponweiser}, \citenamefont {Qiao}, \citenamefont
  {Thöle}, \citenamefont {Tsirkin}, \citenamefont {Wierzbowska}, \citenamefont
  {Marzari}, \citenamefont {Vanderbilt}, \citenamefont {Souza}, \citenamefont
  {Mostofi},\ and\ \citenamefont {Yates}}]{Pizzi_2020}%
  \BibitemOpen
  \bibfield  {author} {\bibinfo {author} {\bibfnamefont {G.}~\bibnamefont
  {Pizzi}}, \bibinfo {author} {\bibfnamefont {V.}~\bibnamefont {Vitale}},
  \bibinfo {author} {\bibfnamefont {R.}~\bibnamefont {Arita}}, \bibinfo
  {author} {\bibfnamefont {S.}~\bibnamefont {Blügel}}, \bibinfo {author}
  {\bibfnamefont {F.}~\bibnamefont {Freimuth}}, \bibinfo {author}
  {\bibfnamefont {G.}~\bibnamefont {G{\'{e}}ranton}}, \bibinfo {author}
  {\bibfnamefont {M.}~\bibnamefont {Gibertini}}, \bibinfo {author}
  {\bibfnamefont {D.}~\bibnamefont {Gresch}}, \bibinfo {author} {\bibfnamefont
  {C.}~\bibnamefont {Johnson}}, \bibinfo {author} {\bibfnamefont
  {T.}~\bibnamefont {Koretsune}}, \bibinfo {author} {\bibfnamefont
  {J.}~\bibnamefont {Iba{\~{n}}ez-Azpiroz}}, \bibinfo {author} {\bibfnamefont
  {H.}~\bibnamefont {Lee}}, \bibinfo {author} {\bibfnamefont {J.-M.}\
  \bibnamefont {Lihm}}, \bibinfo {author} {\bibfnamefont {D.}~\bibnamefont
  {Marchand}}, \bibinfo {author} {\bibfnamefont {A.}~\bibnamefont {Marrazzo}},
  \bibinfo {author} {\bibfnamefont {Y.}~\bibnamefont {Mokrousov}}, \bibinfo
  {author} {\bibfnamefont {J.~I.}\ \bibnamefont {Mustafa}}, \bibinfo {author}
  {\bibfnamefont {Y.}~\bibnamefont {Nohara}}, \bibinfo {author} {\bibfnamefont
  {Y.}~\bibnamefont {Nomura}}, \bibinfo {author} {\bibfnamefont
  {L.}~\bibnamefont {Paulatto}}, \bibinfo {author} {\bibfnamefont
  {S.}~\bibnamefont {Ponc{\'{e}}}}, \bibinfo {author} {\bibfnamefont
  {T.}~\bibnamefont {Ponweiser}}, \bibinfo {author} {\bibfnamefont
  {J.}~\bibnamefont {Qiao}}, \bibinfo {author} {\bibfnamefont {F.}~\bibnamefont
  {Thöle}}, \bibinfo {author} {\bibfnamefont {S.~S.}\ \bibnamefont {Tsirkin}},
  \bibinfo {author} {\bibfnamefont {M.}~\bibnamefont {Wierzbowska}}, \bibinfo
  {author} {\bibfnamefont {N.}~\bibnamefont {Marzari}}, \bibinfo {author}
  {\bibfnamefont {D.}~\bibnamefont {Vanderbilt}}, \bibinfo {author}
  {\bibfnamefont {I.}~\bibnamefont {Souza}}, \bibinfo {author} {\bibfnamefont
  {A.~A.}\ \bibnamefont {Mostofi}}, \ and\ \bibinfo {author} {\bibfnamefont
  {J.~R.}\ \bibnamefont {Yates}},\ }\href@noop {} {\bibfield  {journal}
  {\bibinfo  {journal} {Journal of Physics: Condensed Matter}\ }\textbf
  {\bibinfo {volume} {32}},\ \bibinfo {pages} {165902} (\bibinfo {year}
  {2020})}\BibitemShut {NoStop}%
\bibitem [{\citenamefont {Marzari}\ and\ \citenamefont
  {Vanderbilt}(1997)}]{marzari_97}%
  \BibitemOpen
  \bibfield  {author} {\bibinfo {author} {\bibfnamefont {N.}~\bibnamefont
  {Marzari}}\ and\ \bibinfo {author} {\bibfnamefont {D.}~\bibnamefont
  {Vanderbilt}},\ }\href {\doibase 10.1103/PhysRevB.56.12847} {\bibfield
  {journal} {\bibinfo  {journal} {Phys. Rev. B}\ }\textbf {\bibinfo {volume}
  {56}},\ \bibinfo {pages} {12847} (\bibinfo {year} {1997})}\BibitemShut
  {NoStop}%
\bibitem [{\citenamefont {Li}\ \emph {et~al.}(2020)\citenamefont {Li},
  \citenamefont {Xia}, \citenamefont {Su}, \citenamefont {Yu}, \citenamefont
  {Fu}, \citenamefont {Chen}, \citenamefont {Wang}, \citenamefont {Yu},
  \citenamefont {Zou},\ and\ \citenamefont {Guo}}]{Li}%
  \BibitemOpen
  \bibfield  {author} {\bibinfo {author} {\bibfnamefont {Z.}~\bibnamefont
  {Li}}, \bibinfo {author} {\bibfnamefont {W.}~\bibnamefont {Xia}}, \bibinfo
  {author} {\bibfnamefont {H.}~\bibnamefont {Su}}, \bibinfo {author}
  {\bibfnamefont {Z.}~\bibnamefont {Yu}}, \bibinfo {author} {\bibfnamefont
  {Y.}~\bibnamefont {Fu}}, \bibinfo {author} {\bibfnamefont {L.}~\bibnamefont
  {Chen}}, \bibinfo {author} {\bibfnamefont {X.}~\bibnamefont {Wang}}, \bibinfo
  {author} {\bibfnamefont {N.}~\bibnamefont {Yu}}, \bibinfo {author}
  {\bibfnamefont {Z.}~\bibnamefont {Zou}}, \ and\ \bibinfo {author}
  {\bibfnamefont {Y.}~\bibnamefont {Guo}},\ }\href@noop {} {\bibfield
  {journal} {\bibinfo  {journal} {Scientific reports}\ }\textbf {\bibinfo
  {volume} {10}},\ \bibinfo {pages} {1} (\bibinfo {year} {2020})}\BibitemShut
  {NoStop}%
\bibitem [{\citenamefont {Wu}\ \emph {et~al.}(2018)\citenamefont {Wu},
  \citenamefont {Zhang}, \citenamefont {Song}, \citenamefont {Troyer},\ and\
  \citenamefont {Soluyanov}}]{WU}%
  \BibitemOpen
  \bibfield  {author} {\bibinfo {author} {\bibfnamefont {Q.}~\bibnamefont
  {Wu}}, \bibinfo {author} {\bibfnamefont {S.}~\bibnamefont {Zhang}}, \bibinfo
  {author} {\bibfnamefont {H.-F.}\ \bibnamefont {Song}}, \bibinfo {author}
  {\bibfnamefont {M.}~\bibnamefont {Troyer}}, \ and\ \bibinfo {author}
  {\bibfnamefont {A.~A.}\ \bibnamefont {Soluyanov}},\ }\href {\doibase
  https://doi.org/10.1016/j.cpc.2017.09.033} {\bibfield  {journal} {\bibinfo
  {journal} {Computer Physics Communications}\ }\textbf {\bibinfo {volume}
  {224}},\ \bibinfo {pages} {405 } (\bibinfo {year} {2018})}\BibitemShut
  {NoStop}%
\end{thebibliography}

%
\end{document}